\documentclass[prd,twocolumn,superscriptaddress,altaffilletter,showpacs,nofootinbib]{revtex4}
\usepackage{amssymb}
\usepackage[dvips]{graphicx}
\usepackage{amsmath}

\newcommand{\be}{\begin{equation}}
\newcommand{\ee}{\end{equation}}
\newcommand{\bea}{\begin{eqnarray}}
\newcommand{\eea}{\end{eqnarray}}
\newcommand{\bib}{\bibitem}
\newcommand{\der}{\partial}
\newcommand{\vphi}{\varphi}
\newcommand{\n}{\nabla}

\newcommand{\lab}{\label}

\begin{document}

\title{Conformal transformations and the conformal equivalence principle}

\author{Israel Quiros}\email{iquiros@fisica.ugto.mx}\affiliation{Dpto. F\'isica, Divisi\'on de Ciencias e Ingenier\'ia (DCI) de la Universidad de Guanajuato, A.P. 150, 37150, Le\'on, Guanajuato, M\'exico.}\affiliation{Departamento de Matem\'aticas, Centro Universitario de  Ciencias Ex\'actas e Ingenier\'{\i}as (CUCEI), Corregidora 500 S.R., Universidad de Guadalajara, 44420 Guadalajara, Jalisco, M\'exico.}

\author{Ricardo Garc\'{\i}a-Salcedo}\email{rigarcias@ipn.mx}\affiliation{Centro de Investigacion en Ciencia Aplicada y Tecnologia Avanzada - Legaria
del IPN, M\'exico D.F., M\'exico.}

\author{Jose Edgar Madriz Aguilar}\email{madriz@mdp.edu.ar}\affiliation{Departamento de Matem\'aticas, Centro Universitario de Ciencias Ex\'actas e Ingenier\'{\i}as (CUCEI), Corregidora 500 S.R., Universidad de Guadalajara, 44420 Guadalajara, Jalisco, M\'exico.}

\date{\today}

\begin{abstract}
It is demonstrated that, unless the meaning of conformal transformations for the underlying geometrical structure is discussed on a same footing as it is done for the equations of the given gravity theory, the notion of ``conformal equivalence'', as it has been mostly used in the bibliography to date, lacks physical and mathematical content. A principle of conformal equivalence is then formulated and its consequences are investigated within a conformally invariant modification of Brans-Dicke (BD) theory of gravity. It is also demonstrated that vacuum BD theory ($\omega\neq -3/2$) is not conformally invariant and, consequently, the different conformal frames in which the theory can be formulated are just different -- conformally related -- representations of a same phenomenon. Hence these can not be compared: either both are physically correct, or none is it.
\end{abstract}

\pacs{02.40.-k, 02.40.Ky, 02.40.Hw, 04.20.-q, 04.20.Cv, 04.50.Kd, 04.50.+h, 11.25.Wx}
\maketitle

The importance of invariance of the laws of physics under a units transformation -- also known as conformal transformation -- was recognized long ago by Dicke in Ref.\cite{d-units}. Ironically, publication of this seminal paper signaled the start of an era of confusion regarding the understanding of the (non)equivalence of the different conformally related descriptions of scalar-tensor theories of gravity, among which the Brans-Dicke theory is the prototype \cite{bd}. In BD theory, gravity is mediated by a massless spin-2 metric tensor, together with a (also massless) spin-0 scalar field. Only one of these fields (the metric tensor) is of geometrical origin. This theory represents the most simple generalization of general relativity (GR). It is parametrized by a constant parameter, $\omega$-the BD coupling. In vacuum, the field equations can be derived from the following action:

\be S_{BD}=\frac{1}{16\pi}\int d^4x\sqrt{-g}\;e^\vphi\left[R-\omega (\der\vphi)^2\right],\lab{bd-action}\ee where $R$ is the curvature scalar, $(\der\vphi)^2\equiv g^{\mu\nu}\der_\nu\vphi\der_\mu\vphi$, and we have introduced the scalar field $\vphi$, which is related to the original one in \cite{bd} through, $\phi=e^\vphi$. By means of conformal transformations of the metric, $\bar g_{\mu\nu}=\Omega^2 g_{\mu\nu}$,\footnote{Conformal transformations of the kind discussed here, have nothing to do with general coordinate transformations, so that, for instance, the requirements of conformal invariance and diffeomorphism invariance, would be independent (separate) requirements of a given gravitational theory.} where, $\Omega$-the conformal factor, is a non-vanishing (positive) point-dependent function, this theory can be formulated in different sets of field variables, leading to different formulations of the same theory \cite{d-units}. Amongst them we may cite BD theory in Jordan frame (JF) variables, which is the standard formulation of the theory given in Eq.(\ref{bd-action}) \cite{bd}, and BD theory in Einstein frame (EF) \cite{d-units}. The latter formulation of the theory (for vacuum) can be derived from the action, $\bar S_{BD}=(1/16\pi)\int d^4 x\sqrt{-\bar g}[\bar R-(\omega+3/2)(\bar\der\vphi)^2]$, where the over-bar means the quantities are given in terms of the conformal metric $\bar g_{\mu\nu}$ and the conformal factor has been chosen to be $\Omega^2=e^\vphi$. 

Equivalence of JF and EF formulations of BD theory in particular, and scalar-tensor theories in general, under conformal transformations of the metric, has been discussed in the literature since the early sixties of the last century \cite{d-units}, and more recently has been put to discussion again \cite{de ritis-cqg-1997, faraoni-rev-1999, faraoni-prd-2007}. In spite of the amount of work published on this subject to date the controversy is still open. One of the most evident causes of the debate is the lack of consensus on which meaning to attach to the notion of ``conformal equivalence'', thus rendering the issue a semantic one.

Aim of this letter is to give arguments that might help closing the debate. These arguments are based on the fact that, under conformal transformations of the metric not only the equations of the BD theory are changed but, simultaneously, the underlying geometrical structure of the spacetime manifold is also modified, a fact that is usually dismissed. The change of the geometric properties of the manifold is reflected in that, in terms of the original metric, for instance, the units of the underlying geometry may be point-independent, as it is for Riemannian manifolds, while, in terms of the conformal metric, the units of the geometry may be point-dependent. Although the above fact has been correctly considered in the seminal paper \cite{d-units}, and more recently in \cite{faraoni-prd-2007}, it has passed unnoticed the fact that a geometry with point-dependent/running units can not be Riemannian, but Weyl-integrable (WI) geometry instead. As we will show this point is critical to the understanding of the correct meaning of conformal equivalence. 

Before we pursue the present discussion any further, it will be central to agree on which meaning to attach to the notion of ``conformal equivalence''. We can cite a concrete example where the meaning of the notion of ``equivalence'' is crystal clear: the Einstein's equivalence principle within special relativity (SR-EEP). The physical content of the SR-EEP can be stated in the following simple way: the laws of physics are the same no matter which one of the different inertial reference frames, in which these can be formulated, is chosen. Mathematically this means that there exists a set of linear homogeneous coordinate transformations (Lorentz transformations), that leave invariant the differential equations describing the laws of physics. 

Following the above rule, one can formulate a ``conformal equivalence principle'' (CEP), which might (might not) take place whenever the laws of gravity are involved. From the point of view of its physical content, the CEP can be formulated in the following way: the laws of gravity look the same no matter which one of the different conformally related frames is chosen to describe them. From the mathematical point of view the CEP is to be associated with invariance of the field equations that describe the gravitational phenomena under the conformal transformations of the metric \cite{deser}. In what follows, for definiteness, the statement about conformal equivalence of the different conformal frames in which the laws of gravity can be formulated will entail that the CEP is valid. The contrary statement is also true: if the CEP is not valid, then the different conformally related frames are neither physically nor mathematically equivalent. 

In the present letter, for simplicity, we will focus in vacuum BD-type of gravity theories. Conformal equivalence will entail invariance under the Weyl rescalings/scale transformations,\footnote{Unless it can generate confusion, the terms Weyl rescalings, scale transformations/invariance, and conformal transformations/invariance, will be used interchangeably.}
 
\be \bar g_{\mu\nu}=\Omega^2 g_{\mu\nu},\;\;\bar\vphi=\vphi-2\ln\Omega.\lab{scale-t}\ee Besides, due to the fundamental link between matter and geometry established by general relativity, one might infer that the properties of the units of physics and those of the units of geometry have to be in tight relationship.\footnote{Take, for instance, a measuring stick to which a vector can be attached at a given spacetime point. While the extent of the stick can be the physical unit of measure, the length of its associated vector can be the unit of the underlying geometry.} Consequently, here we shall assume that the physical units of measure and the corresponding units of the geometry share the same mathematical properties. In particular, if the physical units are point-independent, then the units of geometry will be point-independent too. On the contrary, running physical units will imply running units of the underlying geometry. Following this line of reasoning, since the conformal transformation (\ref{scale-t}) links constant units of measure with running ones, one may assume that spacetimes of different geometrical structure stand at the entry and outcome of this transformation, respectively.

At this point we do a step aside to expose the fundamentals of the simplest deformation of Riemann geometry that is able to accommodate running units: WI geometry, which is a subclass in the class of Weyl geometries \cite{w-geom}. A WI space is a manifold ${\cal M}$ endowed with a metric $g_{\mu\nu}$ and a gauge field $\vphi$, such that the ``metricity'' condition, $\n^{(w)}_\mu g_{\alpha\beta}=-\der_\mu\vphi g_{\alpha\beta}$, is fulfilled.\footnote{Arguments found in the literature that point to full mathematical equivalence among Riemann and WI geometry (see, for instance, Ref.\cite{miritzis-cqg}), are questionable. Actually, while the metricity condition and the basic geometric quantities of a WI manifold, are invariant under (\ref{scale-t}), Riemann geometry leaves no room for such invariance property. This is due to the fundamentally constant nature of the measuring units of the Riemann geometry since the metric of a Riemannian manifold is covariantly constant.} Here $\n^{(w)}_\mu$ is the WI-covariant derivative operator, defined in terms of the affine connection of the Weyl-integrable space, $\Gamma^{\;\alpha}_{\beta\gamma}=\{^{\;\alpha}_{\beta\gamma}\}+\left(\delta^\alpha_\beta\der_\gamma\vphi+\delta^\alpha_\gamma\der_\beta\vphi-g_{\beta\gamma}\der^\alpha\vphi\right)/2$, where, as usual, $\{^{\;\alpha}_{\beta\gamma}\}=(1/2)\;g^{\alpha\nu}\left(\der_\beta g_{\nu\gamma}+\der_\gamma g_{\nu\beta}-\der_\nu g_{\beta\gamma}\right)$, are the Christoffel symbols of the metric. For every non-vanishing differentiable function $\Omega$, the affine connection and the metricity condition are invariant under the Weyl rescalings (\ref{scale-t}). Thus the metric, $g_{\mu\nu}$, and the gauge field, $\vphi$, are far from unique: rather these fields belong in an equivalence class of pairs ${\cal C}=\{(g_{\mu\nu},\vphi)|\n^{(w)}_\mu g_{\alpha\beta}=-\n_\mu\vphi g_{\alpha\beta}\}$, such that, any other pair $(\bar g_{\mu\nu},\bar\vphi)$ related with $(g_{\mu\nu},\vphi)$ by a scale transformation (\ref{scale-t}), also belongs in ${\cal C}$. Due to the WI metricity condition, under parallel transport the length of a given vector, $\ell=\sqrt{g_{\mu\nu}\ell^\mu\ell^\nu}$, changes from point to point in the WI manifold, $2 d\ell/\ell=dx^\mu\der_\mu\vphi=d\vphi$. In particular, after parallel transport in a closed path, since, $d\vphi=0$, there is no neat change in the length unit, $d\ell=0$. 

Usually, when a given gravitational action is postulated in the literature, the geometrical structure of the underlying spacetime manifold is implicitly assumed to be Riemannian, unless a different kind of geometry is explicitly stated. It is quite obvious, however, that the kind of geometry which to attach to a given theory of gravity can be only an independent postulate of the theory \cite{d-units}. Consider, as an illustrative example, vacuum BD theory. If one assumes, as it is usually done, that the BD action (\ref{bd-action}) is based on Riemannian spacetimes, then the consequence is that we end-up with a theory of gravity where both the metric and the scalar field propagate the gravity, while only the metric tensor is a geometric object.\footnote{In particular, the affine connection of the Riemann manifold is completely defined in terms of the metric and its derivatives.} The additional scalar field $\vphi$ is non-geometric since $\vphi$ does not participate in the definition of the geometric objects/operators of the Riemann manifold. 

Although it has been claimed in the literature \cite{cho-prl-92} that vacuum BD theory (\ref{bd-action}) is invariant under the scale transformations (\ref{scale-t}), together with the coupling constant redefinition (see also \cite{faraoni-rev-1999,faraoni-omega-infty,appendix}), 

\be \bar\omega=\frac{\omega+6\der_\vphi\ln\Omega(1-\der_\vphi\ln\Omega)}{(1-2\der_\vphi\ln\Omega)^2},\lab{f-red}\ee it can be shown that, in general ($\omega\neq -3/2$), this claim is incorrect. To demonstrate this statement it suffices to show that, even if it is true that the vacuum BD action (\ref{bd-action}) is invariant under (\ref{scale-t}), (\ref{f-red}), the field equations of the theory are not invariant under these transformations, unless $\omega=-3/2$. In fact, since the only conformally invariant vacuum BD-KG equation is necessarily of the form, $\Box\vphi+(\der\vphi)^2/2-R/3=0$, the vacuum BD Klein-Gordon (KG) equation, that can be derived from (\ref{bd-action}),

\be \Box\vphi+\frac{1}{2}(\der\vphi)^2+\frac{R}{2\omega}=0,\lab{v-bd-feqs}\ee is clearly not conformally invariant but for the particular value of the coupling, $\omega=-3/2$. Hence, the conformal invariance of standard (JF) BD gravity theory is, at most, a mirage symmetry. 

The latter result, in conjunction with our statement of the CEP in this letter, leads to conclude that the different conformal representations/frames of vacuum BD theory are neither physically nor mathematically equivalent in general. The fact that the JF and EF formulations of BD theory are mathematically linked through a conformal transformation, $\bar g_{\mu\nu}=e^\vphi g_{\mu\nu}$, does not mean that these representations are equivalent at all (neither physically, nor mathematically).\footnote{Perhaps a closer notion could be ``duality'' rather than ``equivalence'' \cite{quiros}. Duality of the conformal descriptions implies that these are different but mathematically related.} It is obvious, for instance, that the laws of gravity in the JF -- expressed through the JF field equations -- are different from the EFBD gravitational laws. The new ingredient is that the geometrical structure of the spacetime manifold in JFBD also differs from the one in EFBD. In particular, the geodesics over a WI manifold differ from those of the Riemann manifold. Hence, the different conformally related descriptions of the same gravitational phenomenon -- JFBD and EFBD, for instance -- can not be compared: either both are physically correct, or none is it. This result can be easily generalized to scalar-tensor theories in general.

An alternative to vacuum BD gravity theory \cite{bd} could be a theory based on the JFBD action (\ref{bd-action}), but assuming spacetimes of WI geometrical structure instead of Riemannian ones.\footnote{Worth nothing that, in a similar way, one may construct a modification of general relativity consisting of the Einstein-Hilbert action constructed over Weyl-integrable spacetimes.} The resulting theory of gravity would be a truly geometrical theory since both the metric and the scalar field determine the geometric properties of the manifold and, simultaneously, propagate the gravitation. In this case the point-dependent property of the effective gravitational coupling, $G_{eff}\propto e^{-\vphi}$, is intimately linked with the fact that the units of measure of the underlying geometry are running units. However, as it has been already pointed out, unless the coupling parameter, $\omega=-3/2$, this theory is not scale-invariant in spite of the fact that the underlying geometry does. Actually, it is a matter of simple algebra to demonstrate that the particular value of the BD coupling, $\omega=-3/2$, is not transformed by (\ref{f-red}). Hence, the corresponding action, $$S^{BD}_{3/2}=\frac{1}{16\pi}\int d^4x\sqrt{-g}\;e^\vphi\left[R+\frac{3}{2} (\der\vphi)^2\right],$$ together with the field equations derived from it -- in particular Eq.(\ref{v-bd-feqs}) -- will be invariant under the scale transformations (\ref{scale-t}). In other words, the BD action with $\omega=-3/2$ shares invariance under (\ref{scale-t}) with WI spacetimes. It makes sense, then, to re-write the above action in terms of WI quantities by using the Riemannian decomposition of the WI curvature scalar: $R^{(w)}=R-3\Box\vphi-(3/2)(\der\vphi)^2,$ where, in the right-hand-side (RHS) of the above equation, stand usual Riemannian quantities defined in terms of the Christoffel symbols of the metric, while in the LHS there stand WI quantities defined in terms of the WI affine connection instead. The result is,\footnote{We have to point out that in Ref.\cite{salim} a similar action was investigated in a different context.}

\be S^{BD}_{(w)}=\frac{1}{16\pi}\int d^4x\sqrt{-g}\;e^\vphi\;R^{(w)}.\lab{bd-weyl}\ee 

Since the WI Ricci and Einstein's tensors $R^{(w)}_{\mu\nu}$, and $G^{(w)}_{\mu\nu}$, are unchanged by the scale transformations (\ref{scale-t}), in this case not only the action, but also the field equations that can be derived from it, $G^{(w)}_{\mu\nu}=0$, or, in terms of Riemannian quantities,

\bea &&G_{\mu\nu}-\n_\mu\n_\nu\vphi+g_{\mu\nu}\Box\vphi+\frac{1}{2}[\der_\mu\vphi\der_\nu\vphi\nonumber\\
&&+\frac{1}{2}g_{\mu\nu}(\der\vphi)^2]=0,\;\;\Box\vphi+\frac{1}{2}(\der\vphi)^2-\frac{R}{3}=0,\lab{v-wi-bd-feq}\eea are invariant under (\ref{scale-t}). It is evident that the CEP is valid in this formulation of the gravitational laws. 

The resulting scale-invariant modification of BD theory will be a fully geometrical description of the laws of gravity, not sharing any properties with the standard BD theory. Even in the limit, $\vphi\rightarrow\vphi_0$, when Riemann geometry is recovered, the action (\ref{bd-weyl}) is mapped into the standard (Riemannian) Einstein-Hilbert action, $S^{BD}_{(w)}\rightarrow S_{EH}=(1/16\pi G_{eff})\int d^4x\sqrt{-g}\;R$, where the constant, $G_{eff}=e^{-\vphi_0}$, is the effective gravitational coupling, i. e., in that limit general relativity is recovered rather than Brans-Dicke theory.\footnote{General relativity can be obtained from (\ref{bd-weyl}) also by a Weyl rescaling (\ref{scale-t}) with, $\Omega^2=e^\vphi$, since, in this case, WI geometry translates into Riemann geometry spacetimes: $\Gamma^\alpha_{\beta\gamma}\rightarrow\{^{\;\alpha}_{\beta\gamma}\}$, etc. Hence GR is a peculiar member in the class ${\cal C}$, where conformal invariance is a broken symmetry. A Weyl rescaling (\ref{scale-t}) with, $\Omega^2=e^{2\vphi}$, amounts to, $\vphi\rightarrow -\vphi$. Invariance under the latter transformation plays a key role in string-string duality \cite{wands}.}

There are several interesting mathematical consequences arising from the scale-invariant theory of gravity (\ref{bd-weyl}). In particular, due to conformal invariance, instead of a fixed pair $(g_{\mu\nu},\vphi)$ one has a whole -- perhaps infinite -- equivalence class of them ${\cal C}$. The situation is reminiscent of what happens when one invokes invariance under general coordinate transformations: there is not a unique set of coordinates to describe a given physical situation, but a whole (in principle infinite) class of them. While, in the latter case, the spacetime coordinates are meaningless -- the physical meaning is transferred to the invariants of the geometry under spacetime diffeomorphisms -- in the case when conformal invariance is invoked the fields themselves loss independent physical meaning. In this latter case the physically meaningful quantities are the conformal invariants of the WI manifold such as, for instance, the scale-invariant measure of scalar curvature, $e^{-\vphi}\;R^{(w)}$, the scale-invariant measure of spacetime separations, $e^{\vphi/2}ds$, as well as other WI scale-invariant quantities like, $e^{-2\vphi}R^{(w)}_{\mu\nu}R_{(w)}^{\mu\nu}$, etc.

The above discussed scale-invariance property is reflected in the mathematical structure of the field equations: the first and second equations in (\ref{v-wi-bd-feq}) are not independent from each other -- the KG equation is just the trace of the Einstein's one -- so that one equation is redundant. Hence, there will be 6 independent equations to determine 11 unknown degrees of freedom: 10 components of $g_{\mu\nu}$ plus the gauge field $\vphi$. Nonetheless, in addition to the 4 degrees of freedom to make diffeomorphisms (four components of the metric can be transformed away), one more component can be gauged away due to an additional degree of freedom to make scale transformations. I. e., up to general coordinate plus scale transformations, the field equations (\ref{v-wi-bd-feq}) uniquely determine the metric coefficients. The mathematical details, together with the physical implications of this scale-invariant formulation of the laws of gravity, will be discussed in a separate publication (see also the recent paper \cite{romero} where an interesting discussion, very close to ours, is given). 

The results of the present work are important also for the string effective theory. In the first place, the string frame (SF) of the effective dilaton-graviton action, and the corresponding EF are not equivalent. These have to be considered as complementary (yet different) descriptions of a same physical phenomenon: if the SF is depicted by Riemannian geometry, then the EF is described in terms of WI geometry instead. In the second place, since the SF dilaton-gravity action is nothing but JFBD action with $\omega=-1$ \cite{wands},\footnote{This result is independent
of the dimensionality of the spacetime and the number of compactified dimensions.} then, even under the assumption of WI spacetimes, the string effective action can not be conformally invariant, and, in consequence, the CEP is not valid in this case. The consequences of this for string theory will depend on whether the CEP is actually a fundamental principle of nature, as conjectured by Dicke in Ref.\cite{d-units}.

This work was partly supported by CONACyT M\'exico, by the Department of Physics, DCI, Guanajuato University, Campus Le\'on, and by the Department of Mathematics, CUCEI, Guadalajara University.

\end{document}